\documentclass[12pt]{iopart}
\usepackage{graphicx}
\begin{document}
\title{Magnetism and electronic structure calculation of SmN}

\author{C. Morari$^{1}$, F. Beiu\c seanu$^2$, I. Di Marco$^3$, L. Peters$^4$,
E. Burzo$^5$, S Mican$^5$, and L. Chioncel$^{6,7}$}

\address{$^1$ National Institute for Research and Development of Isotopic and
Molecular Technologies, 65-103 Donath,  RO-400293 Cluj Napoca, Romania}
\address{$^2$ Faculty of Science, University of Oradea, RO-410087 Oradea, Romania}
\address{$^3$ Department of Physics and Astronomy, Division of Materials Theory,
Uppsala University, Box 516, SE-75120 Uppsala, Sweden}
\address{$^4$ Institute for Molecules and Materials, Radboud University Nijmegen, NL-6525 AJ Nijmegen, The Netherlands}
\address{$^5$ Babe\c s-Bolyai University Cluj-Napoca, RO-400084 Cluj-Napoca, Romania}
\address{$^6$ Augsburg Center for Innovative Technologies, University of Augsburg, D-86135 Augsburg, Germany}
\address{$^7$ Theoretical Physics III, Center for Electronic Correlations and Magnetism, Institute of Physics,
University of Augsburg, D-86135 Augsburg, Germany}
\ead{liviu.chioncel@physik.uni-augsburg.de}

\begin{abstract}
The results of the electronic structure calculations performed on SmN 
by using the LDA+U method with and without including the spin-orbit coupling are
presented. Within the LDA+U approach, a N(2$p$) band polarization of $\simeq 0.3\ \mu_B$
is induced by Sm(4$f$)-N(2$p$) hybridization, and a half-metallic ground state is obtained. 
By including spin-orbit coupling the magnetic structure was shown to be antiferromagnetic 
of type II, with Sm spin and orbital moments nearly cancelling. This results into a
semiconducting ground state, which is in agreement with experimental results. 
\end{abstract}

\maketitle

\section{Introduction}

Rare-earth nitrides (RN) show a wide range of physical properties,
despite having the same crystal structure  with very close lattice constants 
and similar electronic structures. The structural, magnetic and electronic 
properties of rare-earth monopnictides were reviewed recently~\cite{du.sa.07,na.ru.13}.
In these reviews it was emphasized that the experimental data on the rare-earth monopnictides
are rather scattered, and often lead to contradicting results, due to difficulties in 
obtaining samples of good quality. Thus, controversies concerning their electronic
structure, transport and magnetic properties exist in literature. In the same
reviews~\cite{du.sa.07,na.ru.13} the results of different electronic structure 
calculations of RN compounds were also compared. The description of the strongly 
correlated $4f$ states within the band theory is shown to require methods which
go beyond density functional theory (DFT) in the standard local spin density
approximation (LSDA). 
Band structure calculations were performed on the entire RN series using
the self-interaction correction (SIC) method~\cite{ae.st.04,ho.st.04,sz.te.04},
and the LSDA+U method~\cite{la.la.06,la.la.07}. For the latter, considering 
U corrections for 4$f$ bands only was shown to lead to a half-metallic
ground state~\cite{la.la.07,bu.bu.08}. If also the $5d$ states are corrected
with a local U term, the electronic structure of some RN turns out to
be semiconducting, as shown for example for GdN~\cite{la.la.07}.

SmN was reported to be an antiferromagnet having the N\'eel temperature 
$T_{\rm{N}}$ of 13 K~\cite{sc.wa.66} or 15 K~\cite{bu.ju.65}. From specific 
heat measurements an ordering temperature $T_{\rm{N}}$ of 18.2 K was 
reported \cite{stut.69}. No evidence of magnetic order was shown for 
$T\ge 1.6\ \rm{K}$ by neutron diffraction experiments \cite{mo.ko.79}. 
A magnetic transition around 27 K was reported recently by Preston {\it et al.}
\cite{pr.gr.07}. The rather wide range of the reported ordering temperatures 
can be correlated with the compositions of the samples. The nitrogen vacancy 
lowers the N\'eel temperature, as in the case of GdN \cite{cu.la.75}.
It was argued that SmN is likely to be in fact a metamagnet, however even in 
an applied field of $\mu_0H=6\ \rm{T}$, the low temperature moment was smaller 
than $0.1\ \mu_B$. The transport properties were also studied. A semimetallic 
behaviour of SmN was initially reported \cite{di.go.63}. Later on, SmN was 
shown to be a semiconductor with a band gap of $0.7\ \rm{eV}$ \cite{hull.79}. 
Preston {\it et al.} \cite{pr.gr.07} confirmed experimentally that in the 
studied temperature range ($T\le 150\ \rm{K}$) SmN shows a semiconducting 
behaviour.

Here we report the electronic and magnetic structure of SmN by using an LDA+U+SO approach. 
Several self-consistent solutions for different values of the parameter U exist in both ferro and
anti-ferromagnetic configurations of the samarium moments. We discuss the energy difference
between these self-consistent solutions and show that the ground state structure is that of 
an AF of type II according to Smart's classification \cite{smar.66} with vanishing moments
and semiconducting in agreement with experimental results. 
In our calculations we have used several implementations. The most of our results have been
obtained with the tight-binding linear muffin-tin orbitals (LMTO) method, in the atomic sphere
approximation (ASA), by means of the code LMTO47~\cite{ande.75}. For sake of completeness we
performed several calculations by means of the full-potential (FP) LMTO code
RSPt~\cite{rspt_book,igor1,igor2} and by means of the FP linearized 
augmented plane wave (LAPW) code FLEUR~\cite{fleur_website}. All these implementations
include the on-site Coulomb interaction at the level of LSDA+U~\cite{an.ar.97l}.
The +U contribution on the top of the LSDA represents the additional intra-atomic Hubbard
repulsion among the localized electrons, treated self-consistently in a mean-field (``Hartree-Fock'')
way. Used under appropriate conditions this contribution produces the correct insulating
ground state for several systems where the standard LSDA band theory fails. All implementations
used here consider the most general form of LSDA+U, where the interaction vertex is parametrized
with a full spin and orbital rotationally invariant 4-index U-matrix~\cite{an.ar.97l,sashashick}.
The addition of a Hubbard U interaction term in the energy functional,
also introduces the need for a ``double-counting'' correction. The latter 
accounts for the fact that the Coulomb energy is already included (although
not correctly) in the LSDA functional. The double-counting scheme is unfortunately
not uniquely defined, and usually creates some ambiguity in the LSDA+U
method~\cite{pe.ma.03,yl.pi.09}. Here we adopt the so-called fully localized
limit (FLL) double counting~\cite{an.ar.97l,pe.ma.03,li.an.95} which is suitable
for strongly localized 4f electrons. It is crucial for the present study 
to consider the corrections due to the spin-orbit (SO) coupling self-consistently.
We refer to these calculations in the followings with the acronym LSDA+U+SO~\cite{sh.lu.05}. More
details on the construction of the local orbitals to supplement with  
the Hubbard U term can be found in the references above. Concerning the
strenght of U, instead, we note that a direct evaluation of
the average Coulomb and exchange integrals in terms
of Clebsch-Gordan coefficients and Slater integrals usually produces
unrealistically large values. For this reason Larson {\it et al.} in
Ref.~\cite{la.la.07} scaled the values of the Slater-Coulomb
integrals so that they fit the positions of the 4f-orbitals in Gd pnictides.
The same semi-empirical screening has been applied to SmN, and produced
the values U$=8.22\ \rm{eV}$ and J$=1.07\ \rm{eV}$. 
Given that their study represents the most thourough and systematic
computational work on the RN up to now, for sake of comparison we have
used the same values.

This study is organized in three main parts. In the next section we present the results of LDA+U simulations
by means of the ASA LMTO47 code. Then, the same code is employed to analyze the role of the spin-orbit
coupling. In the last section results from the FP codes are presented, and used as term of comparison
for the ASA results. A brief section summarizing the main conclusions of our study closes this manuscript.

\section{LSDA+U calculations}
\label{lsda+u}
The investigated SmN has the NaCl-structure, with the space group
$Fm\overline{3}m$ (no. 225) in which Sm occupies the $(0,0,0)$
position and N is situated in $(1/2, 1/2, 1/2)$. The computational setup
in ASA codes like LMTO47 involves a minimal set of basis functions for each
atom in the unit cell. The radii of the 
muffin-tin spheres of Sm and N were set to 3.21 a.u. and 2.26 a.u. respectively.
As usual for addressing non-close-packed lattices with ASA codes, empty spheres
are needed. For SmN two empty spheres were introduced in the $(1/4, 1/4, 1/4)$
and $(3/4,3/4,3/4)$ positions. The full Brillouin zone was sampled with a 
Monkhorst-Pack grid of (14 14 14) {\bf{k}}-points. Considering the full
Brillouin zone instead of the irreducible wedge is necessary to identify
all possible local minima in LSDA+U, as discussed below. Finally
the calculations were performed at the 
experimental lattice constant, i.e. $a=5.04\ \AA$, and convergence was considered
up to $\mu$Ry. 

\begin{figure}[h]
\centering
\includegraphics[scale=0.5]{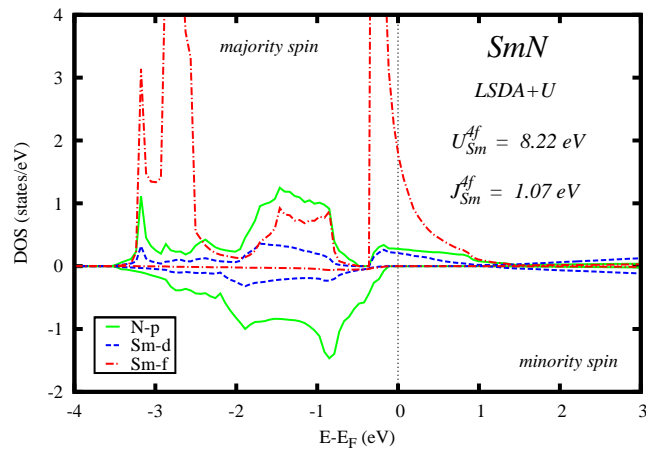}
\caption{LSDA+U orbital resolved density of states of half-metallic SmN. The
minority spin channel shows a gap of $1.04\ \rm{eV}$. Minority spin Sm(4$f$) states are 
situated at higher energies (outside the plot range).} 
\label{fig:SmN}
\end{figure}

The resulting density of states is shown in Figure \ref{fig:SmN}. The majority 
spin channel is metallic, while a gap of about 1.04 {eV} can be seen in 
the minority  spin channel. The $f$-states are formed within the majority spin channel 
in the vicinity of the Fermi level, and at the same time a clear Sm($f$)-N($p$) 
hybridization can also be seen. The Sm($f$)-N($p$) hybridization is less pronounced in the minority 
spin channel. This result is consistent with the electronic 
structure calculation of Svane {\it et al.} \cite{sv.ka.05} that demonstrates the 
presence of $f$-bands at the Fermi level, if the symmetry of the Sm($4f$) density 
matrix is not broken artificially. In this case the three states transforming
according to one of the three dimensional representations of the Oh 
group are partially occupied. If the cubic symmetry is broken, instead, one
can access an insulating solution where one of these states is filled, while
the other two states are empty~\cite{la.la.07}. This solution will be discussed
in the next section, due to that it is the solution obtained when SO coupling
is included.

The N($2p$) states are seen to dominate the top of the valence band in the minority spin channel.
Analysing the band-structure plots (not shown) one notices that the N($2p$) states are separated at the 
X symmetry point by a gap from the Sm($5d$) states which are the dominant character of the bottom 
of the minority spin conduction band. For the majority electrons, where a metallic character is 
evidenced, a strong mixing between Sm($4f$) and Sm($5d$) states and N($2p$) states takes place. 

We note that the X-ray absorption and emission measurements of RN compounds are, in general,
in good agreement with the density of states obtained from LSDA+U calculations, as 
emphasized in Ref.~\cite{na.ru.13}. For SmN, however, our calculations, and also 
those of Ref.~\cite{la.la.07} (with and without broken cubic symmetry), predict
a zero gap in the majority spin band. Furthermore,
there is evidence of some spectral weight at the bottom of the XES
spectra, which was associated with the hybridization between N(2$p$) states and the highest
occupied Sm(4$f$) states in agreement with LSDA+U band structure calculations.
We should also mention that X-ray circular dichroism (XMCD) measurements clearly
demonstrated that the N($2p$) states are magnetically polarized in GdN, so
we could expect that the polarization of N($2p$) is also present in the
SmN compound \cite{lu.pa.05}. This is in fact clearly shown in the results
of Figure \ref{fig:SmN}.

Concerning the magnetic properties, the Sm spin moment is $5.35 \mu_B$, out of which
about $0.05 \mu_B$ belongs to the Sm($5d$) bands, while for N(2$p$) bands a value
of $-0.35 \mu_B$ was determined. The total magnetic moment
per unit cell gives the integer value of $5.00\ \mu_B$, according to the
half-metallic band structure. Obviously, these values cannot be compared with 
experimental data, since magnetism in RN requires the inclusion of SO coupling. For the 
sole purpose of obtaining orbital moments one can resort to the scheme devised by
Larson {\it et al.}~\cite{la.la.07}, where one-additional iteration with SO coupling is run
on top of a converged LSDA+U simulation without SO coupling. In the present study,
however, we cannot follow the strategy of Larson {\it et al.}~\cite{la.la.07}, since it
has not sufficient precision to determine the ground state among several magnetic
configurations. Instead, we will perform full simulations by means of LSDA+U+SO, as
illustrated in the next section. Before presenting these results it is important
to note that the Hubbard U correction on the localized $4f$ states may lead to several
solutions, characterised by a different configuration in the $4f$ local density matrix.
To determine the ground state it is important to allow for all (or at least as many as 
possible) these local minima and identify the global one, as done by 
Larson {\it et al.}~\cite{la.la.07} and more recently by 
Peters {\it et al.}~\cite{lars.pe.89}. 

\section{LSDA+U calculations including spin-orbit coupling}

Neutron diffraction experiments on SmN \cite{mo.ko.79} point to very small
magnetic moments of the order of experimental errors. As a result, these experiments
are not conclusive in determining the ground state magnetic structure, and 
theoretical calculations become of great importance. In this study we have
analyzed theoretically various types of magnetic structures in 
order to obtain information on the ground state. For the actual computation we considered
the rhombohedral representation of the NaCl structure, 
which corresponds to a 4 atom unit cell. In this lattice Sm atoms occupy the positions
Sm1(0,0,0) and Sm2 (1,1,1) and nitrogen atoms are situated at (1/2, 1/2, 1/2) and (-1/2, -1/2, -1/2)
sites. For the simulations with SO coupling, we have performed calculations with and without
empty spheres, and figured out that the latter do not influence the qualitative results
presented in the following. 

\begin{figure}[h]
\centering
\includegraphics[scale=0.3]{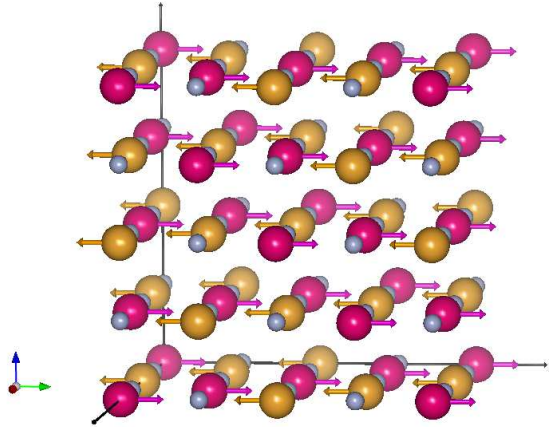}
\caption{(color online) Model for the magnetic structure obtained by non-collinear LSDA+U+SO
calculations. Large red/yellow  circles (arrows) represent Sm1/Sm2 atoms (magnetic moments).
Small gray circles represent the N atoms.}
\label{mag_str}
\end{figure}

Also within the LSDA+U+SO calculations one may obtain several local minima. In the presence 
of the SO coupling the symmetry is lowered and the energy landscape of self-cosnsistent 
solutions is enriched with respect to the number of LSDA+U solutions. 
In the actual calculations, the initial conditions are translated
into the different starting magnetic structures, i.e. the relative orientation
of magnetization axes of the two Sm atoms in the unit cell. This is in fact possible
as the LMTO47 code allows full non-collinear simulations. With this approach
we have considered several ferromagnetic and antiferromagnetic
orderings as starting configurations in which the magnetic moment of
Sm1 atom situated in (0,0,0) was oriented either parallel or anti-parallel 
to the second samarium atom within the unit cell situated at Sm2(1,1,1).
Calculations have been performed with the direction of the moments
along the Cartesian axis. In addition, other initial configurations with the
moments direction along the diagonals of the cube have been also
checked. Further on initial configurations with moments arranged within the 
(XY), (YZ), (ZX),  planes as well as out of plane configurations
in a non-collinear setup have been considered.
The self-consistent total energy calculations with SO coupling were  
performed for a Monkhorst-Pack grid of (24 24 24) {\bf{k}}-points in the full
Brillouin zone. The lowest total energy corresponding to the ground state
magnetic configuration (among those studied) is presented in 
Figure~\ref{mag_str}, and is of type II antiferromagnetic. The energy difference
between the ground-state and the ferromagnetic phase is of about 0.05 Ry for
the selected values of U = 8.22 eV. The type II antiferromagnetic ground state is 
very stable to variations of the strength of U, and was found to have the lowest
energy also for values of U ranging from $6$ to $10\ \rm{eV}$.

\begin{figure}[h]
\centering
\includegraphics[scale=0.1]{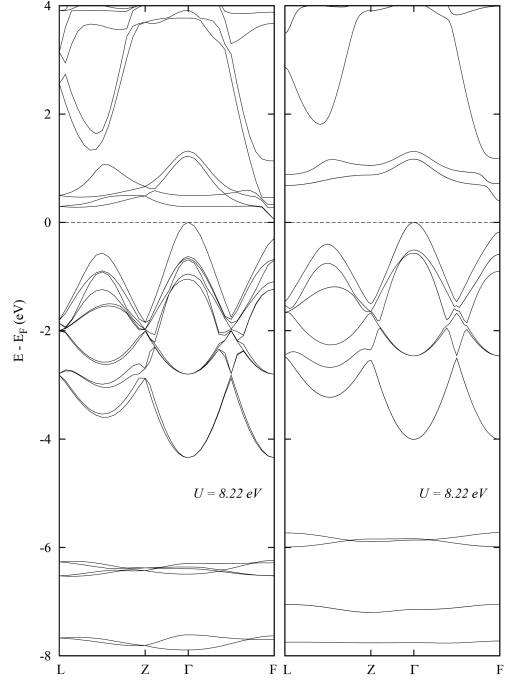}
\caption{(color online) Computed band structures of SmN within the LSDA+U approach including 
spin-orbit coupling for the parallel (left panel) and  anti-parallel (right panel) configurations.}
\label{fig:bnds_SO}
\end{figure}

The computed band structures for the selected values of U and J are presented in 
Figure \ref{fig:bnds_SO}. Both the parallel and anti-parallel arrangements of the 
samarium magnetic moments are presented, corresponding to the ferro and
anti-ferromagnetic ordering. Both configurations exhibit gaps in the band structure. 
However, we must note that
the Sm($5d$) band falls below the Fermi energy at the F point for the parallel
alignment (left panel of Figure \ref{fig:bnds_SO}), so this phase is in 
fact a semimetal, in agreement with Larson {\it et al.}~\cite{la.la.07}.
For the ferromagnetic 
configuration the indirect gap is formed between
the $\Gamma$ point and {\bf k} = $(-0.33 \ 0.00 \ 0.12)$-point, and has 
a size of about 0.01 meV. In the anti-ferromagnetic configuration the gap is formed 
between the $\Gamma$ point and the {\bf k} = $(0.17 \ 0.29 \ 0.12)$-point, and
amounts to 0.40 meV. The size of the indirect gap is particularly significant,
since it can be compared directly with experiment.
Optical absorption measurements indicate the existence of a gap of about 
$0.7\ \rm{eV}$~\cite{hull.79}. We note that we can also simulate the 
non-groundstate anti-ferromagnetic configuration in the LSDA+U approach, 
without SO coupling. In this case we still obtain a semiconductor, but with
a narrower gap of about 0.21 eV. Thus, including SO in a self-consistent way
in addition to the LDA+U approach leads to the correct ground state 
and the right tendency of gap increasing. The magnitude obtained by calculations 
is still underestimated in comparison with the experimental values by a factor of two. 
An important role in the gap opening is played by the 
splitting of the $4f$ manifold within LSDA+U+SO approach. In a cubic crystal field
the $^{6}H_{5/2}$ ground state splits into a $\Gamma_7$ doublet and a
$\Gamma_8$ quartet. Specific heat measurements indicate that the doublet is
lowest in energy. The rombohedral representation used in the computations
has a lower symmetry. This makes it possible to obtain the splitting
described in the previous section without breaking the symmetry with 
ad-hoc corrections through the LSDA+U potential. Results show that the $4f$
states are splitted. The empty $4f$ bands occur about 1 eV above the Fermi
energy, while the occupied $4f$ states are positioned at around 6 eV below
it, for $U=8.22\ \rm{eV}$. This splitting happens for both parallel and
antiparallel alignment, as evident from Figure \ref{fig:bnds_SO}.

From our simulations we can extract the SO coupling constant $\lambda \approx 0.16$ eV. 
This value is significantly smaller than the exchange (Hund) interaction parameter
$J=1.07$ eV, which demonstrates that for this compound a relatively weak SO interaction
exists. This weak interaction results in a Russell-Saunders (LS coupling) scheme 
with the spin $\bi{S}$ and the orbital $\bi{L}$ operators being well defined. In 
such a case the LS-basis is formed by the eigenfunctions of both spin $\bi{S}$
and orbital $\bi{L}$ moment operators. For our parameters U and J, in the
ferromagnetic configuration, 
we obtained a Sm($4f$) contribution to the spin moment equal to 
${\bi{S}}^{Sm} = 5.00 \: \mu_{\rm{B}}$, while the orbital moment
contribution is ${\bi{L}}^{Sm} = - 4.57 \: \mu_{\rm{B}}$. 
In the anti-parallel configuration, instead, we obtained a Sm($4f$) contribution
to the spin moment equal to ${\bi{S}}^{Sm} = 4.99 \: \mu_{\rm{B}}$, 
while the orbital moment
contribution is ${\bi{L}}^{Sm} = - 4.53 \: \mu_{\rm{B}}$. 
In both cases the spin and orbital moments are anti-parallel, i.e. they are
arranged with respect to the third Hund's rule. The atomic-like behavior
of these moments is also clear from their stability with respect
to the magnetic configuration. From the computed values of $|\bi{S}|$ and $|\bi{L}|$ 
the Land\'e factor and the Sm($4f$) magnetic moment were calculated, the obtained 
values being $g=0.18$ and $\mu_{Sm^{f}}=0.36\ \mu_{\rm{B}}$.

It is interesting to analyze how these results change by varying the values of U and J.
In fact, limiting our calculations to values extracted to match a few selected experimental
properties is too much prone to an error, especially in light of the limitations that
experiment have for RNs. We have therefore evaluated the magnetic properties for a higher
value of U, i.e. U$=9.22$ eV. This corresponds to an increment of 1 eV with respect to our
selected value. We obtained a SO coupling constant $\lambda \approx 0.17$ eV. The magnetic
moments due to the Sm($4f$) states in the anti-parallel configurations became
${\bi{S}}^{Sm} = 4.97 \: \mu_{\rm{B}}$ and ${\bi{L}}^{Sm} = - 4.63 \: \mu_{\rm{B}}$. These 
values exhibit an expected increase of the orbital polarization, which is compensated by a
slight decrease of the spin moment. The computed Land\'e factor and Sm($4f$) magnetic moment 
were $g=0.21$ and $\mu_{Sm^{f}}=0.55\ \mu_{\rm{B}}$. These values lead to the same
physical picture outlined above.

Thus, we conclude that the magnetic structure of SmN is essentially of AF-type II. Such magnetic 
structures were shown in GdX \cite{li.ha.97} and TbX \cite{hull.79}, with X = P, As, Sb and Bi.
Although the spin and orbital moments of Sm nearly cancel, finite values of SmN magnetizations
can be obtained. These can be correlated with the contributions of $Sm(5d)$ and N(2$p$) 
band polarizations resulting mainly from the hybridization with the Sm(4$f$) band. 

\section{Full potential calculations}
Given that the structure of SmN is not strictly close-packed, we have checked our main
results by means of the RSPt code~\cite{rspt_book,igor1,igor2}, which is based
on the FP-LMTO method. The calculations were limited to the AFM phase
of SmN, and were performed with $Sm(6s,6p,5d,4f)$ and $N(2s,2p)$ 
orbitals for the valence electrons. The radii of the muffin-tin spheres of
Sm and N were set to 2.5 a.u. and 2.0 a.u. respectively. The number of kinetic
energy tails, describing the basis in the interstitial region 
between the muffin-tin spheres, was set to 3 for the $s$ and $p$ states,
and to 2 for the $d$ and $f$ states. 
These parameters have been carefully checked to offer a converged basis. 
The full Brillouin zone was sampled with a Monkhorst-Pack grid of 
(24 24 24) {\bf{k}}-points. 

\begin{figure}[h]
\centering
\includegraphics[scale=0.5]{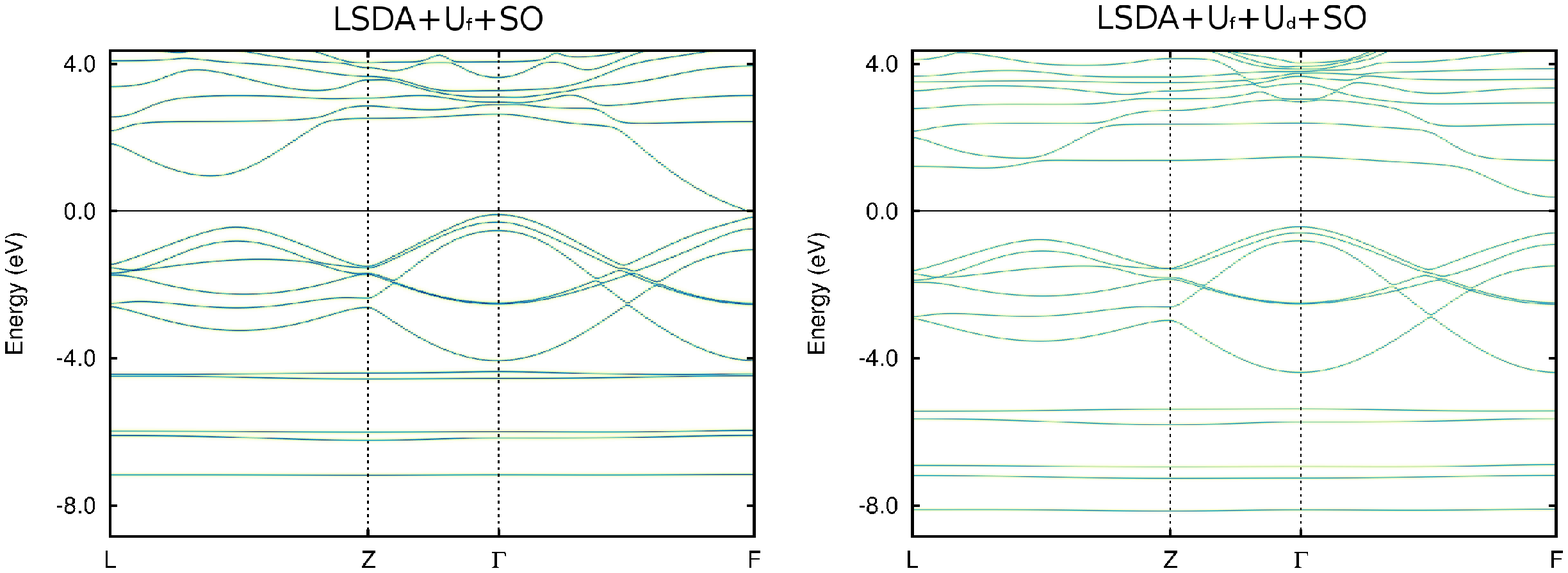}
\caption{(color online) Computed band structures of SmN within the LSDA+U 
approach including spin-orbit coupling by means of the FP-LMTO code RSPt. Results
without and with an additional Hubbard correction U$_d$ are shown, respectively
in the left panel and right panel.}
\label{fig:fp_res}
\end{figure}

We have analyzed the magnetic anisotropy in the AFM phase of type II. We have
confirmed that the easy axis is aligned towards the y-direction, for both
DFT in LDA with SO coupling and LSDA+U+SO. For the simulations where the 
magnetization is aligned along the easy axis, the values of
the Sm($4f$) contribution to the spin and orbital moment  are
${\bi{S}}^{Sm} = 4.09 \: \mu_{\rm{B}}$ and
${\bi{L}}^{Sm} = - 3.46 \: \mu_{\rm{B}}$. Although to a first sight these 
values may look rather different than those presented in the previous
section, we have to keep in mind that in a FP code the muffin-tin spheres 
cannot overlap, and an important contribution to the magnetic moments may
lay in the interstitial region. In addition to the magnetic properties,
we have also computed the electronic structure, which is reported in
the left panel of Figure \ref{fig:fp_res}. Here one can see a major 
difference with respect to the data reported in Figure \ref{fig:bnds_SO}. 
In fact the electronic structure obtained in a FP code presents
semi-metallic character, similarly to what observed above for the FM
phase. The presence of states within the gap changes considerably 
the physical picture related to the band structure with
respect to the LMTO47 simulations. To understand if this behavior 
is due to the ASA character or to more implementation specific details,
we performed additional FP simulations with the LAPW code 
FLEUR~\cite{fleur_website}. Indeed, we obtained very similar band
structures (not shown) as those reported in the left panel
of Figure \ref{fig:fp_res}. 
The difference among ASA and FP calculations creates an important problem.
On the one hand the FP simulations must be preferred for accuracy and
reliability, on the other hand the experimental data seem to support
the insulating solution obtained in ASA (see above). The problem of 
the absence of an energy gap has been discussed in detail
by Larson {\it{et al.}}~\cite{la.la.07} for the ferromagnetic phase.
There the authors suggest that an insulating character can be 
obtained by means of an additional Hubbard correction
for the Sm($5d$) states. They determine a U parameter equal to 6.4 eV
in order to fix the experimental band gap of GdN, while J was set to 
zero for sake of simplicity. With these parameters they obtain a
semi-metallic solution for the FM phase of SmN, where the indirect
band gap is of about -0.15 eV. We have followed their strategy, and
performed LSDA+U+SO simulations of the AFM phase with this setup. 
The resulting band
structure is reported in the right panel of Figure \ref{fig:fp_res}.
A band gap is now opened, analogously to what reported in 
Figure \ref{fig:bnds_SO}. It is interesting to note how the 
Hubbard correction to the Sm($5d$) states also changes the position
(but not the relative splitting) of the Sm($4f$) levels.

\section{Conclusion}
In the present paper we have discussed the magnetic and the electronic
structure properties of the SmN compound. The sensitivity of the macroscopic
properties to temperature, pressure and impurities
is primarily due to the complex interplay between the rare-earth
$4f$-$5d$ interactions and the hybridization 
between $Sm(4f)$ states with the $N(2p)$ states. The electronic structure 
results based on the LSDA+U method favor a half-metallic ground state, whereas
if we include the spin-orbit coupling a magnetic AF-type II 
structure is obtained with moments that nearly cancel. 
The nearly linear field dependence of the SmN magnetization, at 2 K, 
experimentally observed in a thin film close to being
stoichiometric, \cite{pr.gr.07} is not in contradiction with AF ordering.
In addition, the presence of SO coupling  leads to and enlarged 
semiconducting gap. However, the magnitude of the gap is still
underestimated, in comparison with the experimental measurements, 
an indication that the present LDA+U+SO calculations
does not give a complete description. Full-potential calculations
show, in fact, that the observed gap arises mainly due to 
the restrictions imposed by the atomic-sphere approximation
and not because of the Coulomb interaction. A more physical
description of the band gap can be obtained with an explicit
Hubbard correction for the Sm($5d$) states. However, this
approach presents two important problems. First, it 
introduces another parameter in the simulations; second
it is indeed hard to justify the presence of a local 
Hubbard term to improve the description of the highly
delocalized $5d$ states. In this context, ASA simulations
have the advantage of giving a physically correct solution
without needing additional semi-empirical corrections. Anyway,
once the band gap is present, possible metal-semiconductor
transitions are expected upon doping or upon changing
external conditions. In such cases an important role is 
played by not only the aforementioned $d$-$d$ interaction, but
also by the $f$-$d$ and $f$-$s$ hybridizations. These effects
are finally entangled with the multiplet structure of the 
atomic-like $4f$ states. In order to consider the multiplet 
structure properly, an approach beyond the static mean-field
is needed. The LDA+DMFT approach~\cite{gabi.06,karsten.07,li.ka.98}
in the Hubbard I approximation~\cite{li.ka.98,hubb.63} is capable
of capturing all these effects, and has already been applied
to the paramagnetic ErAs~\cite{po.de.09}, and to TbN~\cite{lars.pe.89}.
This type of calculations introduce additional
difficulties that we plan to address in our future research.

\ack 
The calculations were performed in the Datacenter of NIRDIMT. 
C Morari acknowledges the financial support offered by the Augsburg 
Center for Innovative Technologies (ACIT), University of Augsburg,
Germany. S Mican would like to acknowledge support from project
POSDRU/88/1.5/S/60185-``Innovative doctoral 
studies in a knowledge based society''. I Di Marco and L Peters
acknowledge the Nederlandse Organisatie voor Wetenschappelijk
Onderzoek (NWO), and the Swedish National Allocations Committee
(SNIC/SNAC) for computational time
at the National Supercomputer Cluster (NSC).
Also SURFsara is acknowledged for the usage of LISA
and their support.


\section*{References}
\begin{thebibliography}{27}
\bibliographystyle{unsrt}
\bibitem{du.sa.07} Duan C-G, Sabirianov R F, Mei W N, Dowben P A, Jaswal S S and Tsymba E Y 2007 {\it \JPCM} {\bf 19} 315220
\bibitem{na.ru.13} Natali F, Ruck B J, Plank N O V, Trodahl H J, Granville S, Meyer C and Lambrecht W R L 2013 {\it Progress in Materials Science} {\bf 58} 1316
\bibitem{ae.st.04} Aerts C M, Strange P, Horne M, Temmerman W M, Szotek Z and Svane A 2004 {\it \PR}B {\bf 69} 045115
\bibitem{ho.st.04} Horne M, Strange P, Temmerman W M, Szotek Z, Svane A and Winter H 2004 {\it \JPCM} {\bf 16} 5061--70
\bibitem{sz.te.04} Szotek Z, Temmerman W M, Svane A, Petit L, Strange P, Stocks G M, K\"{o}dderitzsch D K, Hergert W and Winter H 2004 {\it \JPCM} {\bf 16} S5587-S5600
\bibitem{la.la.06} Larson P and Lambrecht W R L 2006 {\it \PR}B {\bf 74} 085108
\bibitem{la.la.07} Larson P, Lambrecht W R L, Chantis A and Van Schilfgaarde M 2007 {\it \PR}B {\bf 75} 045114

\bibitem{bu.bu.08} Burzo E, Bucur N, Allmaier H and Chioncel L 2008 {\it J. Opt. Adv. Mat} {\bf 10} 389
\bibitem{sc.wa.66} Schumacher D P and Wallace W E 1966 {\it Inorg. Chem.} {\bf 5} 1563 
\bibitem{bu.ju.65} Busch G, Junot P, Levy F, Menth A and Vogt O 1965 {\it Phys. Lett.} {\bf 14} 264
\bibitem{stut.69} Stutius W 1969 {\it Phys. Kondens. Matt.} {\bf 10} 152
\bibitem{mo.ko.79} Moon R M and Koehler W C 1969 {\it \JMMM} {\bf 14} 256
\bibitem{pr.gr.07} Preston A R H, Granville S, Housden D H, Ludbrook B, Ruck B J, Trodahl H J, Bittar A, Williams G V M, Downes J E and DeMasi A \etal 2007 {\it \PR} B {\bf 76} 245120
\bibitem{cu.la.75} Cutler R A and Lawson A W 1975 {\it \JAP} {\bf 46} 2739
\bibitem{di.go.63} Didchenko D and Gortsema F P 1963 {\it J. Phys. Chem. Solids} {\bf 24} 863
\bibitem{hull.79} Hullinger F 1979 {\it Handbook on the Physics and Chemistry of Rare Earths} vol~4 ed Gschneider K A and LeRoy E (Amsterdam: North-Holland) pp~153--256

\bibitem{smar.66} Smart J S 1966 {\it Effective Field Theories of Magnetism} (Philadelphia: Sanders)
\bibitem{ande.75} Andersen O K 1975 {\it \PR} B {\bf 12} 3060--83

\bibitem{rspt_book} Wills J M, Alouani M, Andersson P, Delin A, Eriksson O and Grechnyev O 2010 {\it Full-Potential Electronic Structure Method} in {Electronic Structure and Physical Properties of Solids: Springer Series in Solid-State Sciences} (Berlin: Springer-Verlag)
\bibitem{igor1} Di Marco I, Min\'ar J, Chadov S, Katsnelson M I, Ebert H and Lichtenstein A I 2009 {\it \PR}B {\bf 79} 115111
\bibitem{igor2} Gr{\aa}n\"as O, {Di Marco} I, Thunstr\"om P, Nordstr\"om L, Eriksson O, Bj\"orkman T and Wills J M 2012 {\it Computational Materials Science}  {\bf 55} 295
\bibitem{fleur_website} {\it http://www.fleur.de}
\bibitem{an.ar.97l} Anisimov V I, Aryasetiawan F and Lichtenstein A I 1997 {\it \JPCM} {\bf 9} 767--808
\bibitem{sashashick} Shick A B, Drchal V and Havela L 2005 {\it Europhys. Lett.} {\bf 69} 588
\bibitem{pe.ma.03} Petukhov A G, Mazin I I, Chioncel L and Lichtenstein A I 2003 {\it \PR}B {\bf 67} 153106
\bibitem{yl.pi.09} Ylvisaker E R, Pickett W E and Koepernik K 2009  {\it \PR}B {\bf 79} 035103
\bibitem{li.an.95} Liechtenstein A I, Anisimov V I and Zaanen J 1995 {\it \PR}B {\bf 52} R5467
\bibitem{sh.lu.05} Shorikov A O, Lukoyanov A V, Korotin M A and Anisimov V I 2005 {\it \PR}B {\bf 72} 024458
\bibitem{lars.pe.89} Peters L, {Di Marco} I, Thunstr\"om P, Katsnelson M I, Kirilyuk A and Eriksson O 2014 {\it \PR}B {\bf 89}, 205109
\bibitem{sv.ka.05} Svane A, Kanchana V, Vaitheeswaran G, Santi G, Temmerman W M, Szotek Z, Strange P, and Petit L 2005 {\it \PR}B {\bf 71} 045119


\bibitem{lu.pa.05} Leuenberger F, Parge A, Felsch W, Fauth K, and Hessler M 2005 {\it \PR}B {\bf 72} 014427
\bibitem{li.ha.97} Li D X, Haga Y, Shida H, Suzuki T, Kwon Y S and Kido G 1997 {\it \JPCM} {\bf 9} 10777

\bibitem{gabi.06} Kotliar G, Savrasov S Y, Haule K, Oudovenko V S, Parcollet O, and Marianetti C A 2006 {\it Review Modern Physics} {\bf 78} 865
\bibitem{karsten.07} Held K 2007 {\it Advances in Physics} {\bf 56} 829
\bibitem{li.ka.98} Lichtenstein A I, and Katsnelson M I 1998 {\it \PR}B {\bf 57} 6884--95
\bibitem{hubb.63} Hubbard J 1963 {\it Proc. R. Soc. London} vol~276 p~238
\bibitem{po.de.09} Pourovskii L V, Delaney K T, Van de Walle C G, Spaldin N A and Georges A 2009 {\it \PRL} {\bf 102} 096401

\end{thebibliography}
\end{document}